\documentclass[aps,prb,twocolumn,showkeys,amssymb,amsmath,nobibnotes,nofootinbib,superscriptaddress]{revtex4}
\usepackage[colorlinks=true,linkcolor=red]{hyperref}
\usepackage{amsmath}
\usepackage{amsfonts}
\usepackage{amssymb}
\usepackage{verbatim}
\usepackage{times}
\usepackage{color}

\usepackage{graphicx}
\usepackage[sort&compress]{natbib}

\newcommand{\braket}[2]{\langle #1| #2 \rangle}
\newcommand{\bra}[1]{\langle #1|}
\newcommand{\ket}[1]{|#1\rangle}
\newcommand{\Tr}[1]{{\mathrm{Tr}}\left(#1\right)}
\newcommand{\avg}[1]{\langle #1\rangle}
\newcommand{\Ord}[1]{{\mathcal O}\!\left(#1\right)}
\newcommand{\mean}[1]{\overline{#1}}
\def\cS{{\mathcal S}}

\def\cI{{\mathcal I}}
\def\cC{{\mathcal C}}
\def\id{{\mathbb I}}

\begin{document}


\title{On the structure of typical states of a disordered Richardson model and many-body localization}



\author{F.Buccheri}
\author{A.De Luca}
\affiliation{SISSA - 
via Bonomea 265, 34136, Trieste, Italy}
\affiliation{INFN, Sezione di Trieste -
via Valerio 2, 34127 Trieste, Italy
}
\author{A.Scardicchio}
\affiliation{INFN, Sezione di Trieste -
via Valerio 2, 34127 Trieste, Italy
}
\affiliation{Abdus Salam ICTP - 
Strada Costiera 11, 34121, Trieste, Italy}



\date{\today}

\begin{abstract}
We present a thorough numerical study of the Richardson model with quenched disorder (a fully-connected XX-model with longitudinal random fields). We find that for any $g>0$ the eigenstates occupy an exponential number of sites on the unperturbed Fock space but that single-spin observables do not thermalize, as tested by a microcanonical version of the Edwards-Anderson order parameter $q>0$. We therefore do not observe MBL in this model. We find a relation between the inverse participation ratio, $q$ and the average Hamming distance $L$ between spin configurations covered by a typical eigenstate for which we conjecture a remarkably simple form for the thermodynamic limit $L/N=\frac{g}{2(1+g)}$. We also studied the random process defined by the spread of a typical eigenstate on configuration space, highlighting several similarities with hopping on percolated hypercube, a process used to mimic the slow relaxation of spin glasses. 
A nearby non-integrable model is also considered where delocalization is instead observed, although the presence of a phase transition at infinite temperature is questionable.

\end{abstract}


\maketitle

\section{Introduction}

Recently \cite{basko2006metal,basko2006problem} it has been pointed out that the phenomenon of Anderson localization \cite{anderson1958absence}, usually associated with single-particle hopping in a random potential, can be present even in the many-body eigenstates of an interacting quantum system and manifest itself as a phase transition at finite and even infinite temperature. 

This phenomenon has been dubbed \emph{many-body localization} (henceforth MBL) and it can be conceived as an example of AL on configuration space, rather than real space. As the geometry of configuration space for a many-body system is quite different from that of a regular lattice in few dimensions, MBL is thought to have properties distinct from those of the single-particle AL.

MBL should be responsible, among other things, of the exact vanishing of the DC conductivity of metals below a critical temperature \cite{basko2006metal} and of the failure \cite{altshuler2010anderson, knysh2010relevance} of the simplest version (and possibly of all versions) of the quantum adiabatic algorithm \cite{farhi2001quantum} for the solutions of NP-complete problems; it has also been studied in disordered Heisenberg spin-chains \cite{brown2008quantum, pal2010many-body} where the phase transition has been linked to the infinite-randomness fixed point. The similarity of some features of MBL to the glass transition in spin and configurational glasses makes it the closest to a \emph{quantum} analog of a glass transition, where the assumptions of equilibrium statistical mechanics fail.

As we said, in some problems MBL is found in typical many-body states\cite{oganesyan2007localization}, namely states sampled with uniform distribution from the spectrum (therefore corresponding to infinite temperature). These states are difficult to study directly, much more than the ground states for which many approximations (DMRG, MPS etc.) can be devised: indeed the only strategy here seems to be exact diagonalization (as used in \cite{pal2010many-body} for example), the exponential complexity of which limits the size of the systems to less than 20 spins; alternatively the study of correlation functions with time-dependent DMRG was used, whose failure to converge due to growing entanglement can signal the onset of delocalization\cite{znidaric2008many}. Analytic results have mainly been obtained by studying the behavior of the perturbation theory for increasing system sizes \cite{knysh2010relevance}. 

In this work we report the results of our numerical study on the structure of typical states of a fully-connected quantum spin model with quenched disorder (the Richardson model \cite{richardson1963restricted, richardson1964exact, dukelsky2004colloquium}) which has been introduced as a model of nuclear matter and has been studied in connection with the finite-size scaling of the BCS theory of superconductivity. Integrability allows us to go to sensibly higher spin numbers ($N=50$ spins for single states and we will collect extensive statistics up to $N=40$) and therefore make some educated guesses on the thermodynamic limit of the system. 

We will begin by discussing the method we devised for the solution of the Bethe-ansatz equations (the Richardson equations) which is at variance with respect to those very refined ones, used in the study of ground state and low-lying excited states \cite{faribault2009bethe, faribault2008exact} (our method will be close in spirit to that used in \cite{faribault2011numeric}, which appeared while this work was being completed).
Then we will discuss the observables, since we will face the problem that the classical observables in localization studies, the inverse participation ratio (IPR), is computationally heavy (its complexity goes as $2^N$, although still smaller than $\Ord{2^{3N}}$ steps required by exact diagonalization).
We devised a Montecarlo method for the measure of IPR and performed an extensive study of an Edwards-Anderson order parameter $q$ \cite{mezard1987spin}, which is related to the average (1,$N$-1)-entanglement (the Meyer-Wallach\cite{meyer2002global} entanglement measure) and to the average Hamming distance $L$ between states in the computational basis whose superposition forms the eigenstate. As we show, $q$ is related to the long-time spin relaxation and therefore thermalization is not achieved as long as $q>0$.

We also observe that for the Richardson model, $q$ is in 1-to-1 correspondence with the IPR $\cI$ (but the relation is different from what found for disordered Heisenberg model\cite{viola2007generalized}). We find for the thermodynamic limit of $q$ the deceptively simple expression as a function of the hopping $g$: $q=(1+g)^{-1}$ which implies in the same limit, for the average Hamming distance $L/N=\frac{g}{2(1+g)}$. We will define a local entropy density $s=\log\cI/2L$, for which we find numerically a well-defined thermodynamic limit, although the limiting form does not seem to have the simple character of the previous quantities. 

We have also studied the clustering properties of the eigenstates, and we have not found any presence of clusters but rather as the hopping $g$ is increased the eigenstates spread rather uniformly over the configuration space. Neighboring states in energy have very close values of $q$ but their overlap (a measure common to spin-glass studies) is close to 0. This means that clusters can be formed if one takes a superposition of states in a small energy interval to make a microcanonic density  matrix.

The picture that emerges from this analysis is that there is no many-body localization-delocalization phase transition in this model{\color{blue}:} although the states appear de-localized on the computational basis for any finite $g$, the average single-spin observables are always localized. 

Finally, we discuss the role of integrability in the previous predictions and the implications of our findings for more natural cases where integrability is broken\cite{canovi2010quantum}.

\section{The model and its solution} 

The Richardson model \cite{richardson1963restricted, richardson1964exact} is an XX-model (i.e.\ with no $s^zs^z$ coupling) of pairwise interacting spins with arbitrary longitudinal fields
\begin{equation}\label{hamiltonian}
H=-\frac{g}{N}\sum_{\alpha,\beta=1}^N s^+_\alpha s^-_\beta-\sum_{\alpha=1}^N h_\alpha s^z_\alpha,
\end{equation}
where $s^{x,y,z}$ are spin-$\frac{1}{2}$ representation of $SU(2)$ algebra. This model can accommodate quenched randomness in the arbitrary choice of the fields $h_\alpha$.\footnote{It is known that this model in 1-d reduces to non-interacting fermions and hence localizes for arbitrary small disorder. This is a peculiarity of 1-d and is due to the existence of the Jordan-Wigner map.} We choose a Gaussian distribution for them, with $\mean{h}=0, \mean{h^2}=1$. First of all one notices that the total spin $S^z$ is conserved and we focus on the subspace $S_z=0$ which exists only for even $N$. All the states in the sector $S^z=(2M-N)/2$ can be found by applications of $M$ generalized raising operators on the state with all spin down:
\begin{equation}\label{eigenstates}
\ket{E[w]}= \prod_{j=1}^M B(w_j) |\downarrow\ldots\downarrow\rangle,
\end{equation}
where the $M$ \emph{Richardson roots} $w_j$ satisfy the $M$ \emph{Richardson equations}:
\begin{equation}\label{RE}
\forall j=1,...,M:\quad
\frac{N}{g}+\sum_{\alpha=1}^N\frac{1}{w_j-h_\alpha}-\sum_{k=1, k\neq j}^M\frac{2}{w_j-w_k}=0
\end{equation}
in terms of which the raising operators are
\begin{equation}\label{B}
B(w)=\sum_{\alpha=1}^N\frac{S_\alpha^+}{w-h_\alpha} 
\end{equation}
and the energy of the state is given by
\begin{equation}\label{nrg}
 E[w]=\sum_{j=1}^M w_j-\sum_{\alpha=1}^N \frac{h_\alpha}{2}.
\end{equation}
As we said, we will focus on $S_z=0$ so we will have $M=N/2$, which means that we have to solve $N/2$ coupled nonlinear equations, the different $\binom{N}{N/2}$ states are determined by the boundary conditions as $g\to 0$. Indeed as $g\to 0$ the roots tend to some of the fields $h_\alpha$ and the choice of the set can be used to label the state at any $g$.

Starting from $g=0$ instead, and increasing $g$, the roots start departing from their initial $h$'s values, collide and become complex conjugate. By increasing $g$ sometime they recombine and return real. The number of roots that eventually ($g\to\infty$) diverge is equal to the total spin $S$ of the state (which is a conserved quantum number at infinite $g$). An algorithm which can follow the evolution of the roots with $g$ has to take into account these changes in the nature of the solution, where the roots become complex conjugate. These critical points, for random choices of the $h$'s can occur at particularly close values of $g$ and this can create troubles for the algorithm.\footnote{This problem is not so serious for the ground state and first excited states so one can go to much higher values of $N$ without losing accuracy.} In order to pass the critical points a change of variable is needed, and one natural choice is \cite{faribault2009bethe}:
\begin{eqnarray*}
 && w_+ = 2 h_c-w_1-w_2  \\
 && w_- = (w_1-w_2)^2
\end{eqnarray*}
in which $w_1$ and $w_2$ are the root colliding on the level $h_c$.

When more than a pair of roots collide in a too small interval of $g$ this change of variables may not be sufficiently accurate and one should think of something else (if one does not want to reduce the step in the increment of $g$ indefinitely). The most general change of variables which smooths out the evolution across critical points is that which goes from the roots $w_j$ to the coefficients $c_i$ of the characteristic polynomial $p(w)$ --i.e.\ the polynomial whose all and only roots are the $w_j$'s.

This polynomial is quite interesting in itself as it satisfies a second order differential equation whose polynomial solutions have been classified by Heines and Stjielties \cite{szego1939orthogonal}.\footnote{The equation is $-h(x)p''(x)+\left(\frac{h(x)}{g}+h'(x)\right)p'(x)-V(x)p(x)=0,$ where $h(x)=\prod_{\alpha=1}^N(x-h_\alpha)$, $V(x)=\sum_{\alpha=1}^N\frac{h(x)A_\alpha}{x-h_\alpha}$.  The problem to be solved is to find a set of $A_\alpha$'s such that there exists a polynomial solution of this equation. The solution will automatically satisfy also $A_\alpha=\frac{p'(h_\alpha)}{p(h_\alpha)}.$ A similar approach has also been investigated in the recent work \cite{faribault2011numeric}.}  Following the evolution of the coefficients $c_i(g)$ is a viable alternative to following the roots but we found out that the best strategy is a combination of both evolutions. Therefore we follow the evolution of the roots, extrapolating the coefficients and using them to correct the position of the roots at the next step in the evolution. In this way we do not implement any change of variables explicitly and we do not have to track the position of critical points. This algorithm
 \footnote{Python code is available on the webpage: \\
http://www.sissa.it/statistical/PapersCode/Richardson/} can be used on a desktop computer to find the roots of typical states with about 50 spins, although in order to collect extensive statistics we have limited ourselves to $N=40$. 

\section{Order parameters}
\subsection{IPR, entanglement, average Hamming radius of an eigenstate and a local entropy}
Once obtained the roots one is faced with the task of studying the state. The quantity characterizing the localization/delocalization properties of a state (on the basis of $s^z_i$'s: the computational basis or configuration space $\cC$) is the inverse participation ratio of an eigenstate $\ket{E}$:
\begin{equation}
\cI=\left(\sum_{{s_1,...,s_N,\ \sum_\alpha s_\alpha=0}}|\braket{s_1,...,s_N}{E}|^4\right)^{-1}.
\end{equation}
We will see that for all $g>0$, $\log\cI\propto N$, i.e., an exponential number of sites of the hypercube of spin configurations is covered;
nevertheless, we generally observe that
\begin{equation}
\lim_{N\to\infty} \frac{\cI}{\binom{N}{N/2}}\to 0,
\end{equation}
which flags instead a single-particle \emph{localized} phase, according to the definition common in AL studies. In fact, the analysis of single-particle observables will confirm this scenario.
 
The amplitudes $\braket{s_1,...,s_N}{E}$ can be calculated as ratio of determinants of $(N/2)\times (N/2)$ matrices (therefore in time $\sim N^3$) once the roots $w_j$ are known. However the number of terms in the sum is exponential in $N$ so the calculation of $\cI$ requires an exponential number of terms\footnote{We have looked for a shortcut to evaluate this sum but to our knowledge integrability does not help us here.} and we are limited again to twenty spins or so.

We found two ways around this difficulty: they are complementary and can be checked one against the other for consistency. First, we devised a Montecarlo algorithm for the evaluation of $\cI$. Define the probabilities $p_a=|\braket{a}{E}|^2$ where $a\in\cC$ stands for one of the $\binom{N}{N/2}$ allowed configurations of spins which constitute the configuration space $\cC$. We perform a random walk with the probabilities $p_a$'s, namely start from a random configuration $a$ and we try to move to a random one of the $(N/2)^2$ neighboring states, say $b$, by accepting the move with probability $\min(1,p_b/p_a)$. This involves only one computation of $p_b$, which takes time $\sim N^3$. The random walk proceeds in this way, generating a history of configurations $a$ for which we can take the average over Montecarlo time of $p_a$. The inverse of this value gives $\cI$. 

The intensive quantity is $\log\cI /N$, which can then be averaged over different states and realizations. We observe that for all $g=\Ord{1}$ the value of $\ln\cI\propto N$, testifying then that each state occupies an exponential number of states in the configuration space. 

The second method is to find another quantity which can be computed in polynomial time and to link it to $\cI$. Since the average values $\bra{E}s_\alpha^z\ket{E}$ can be expressed again in terms of determinants they can be calculated in $\Ord{N^3}$ time: therefore one is led to consider a \emph{microcanonical} version the Edwards-Anderson order parameter associated to a single eigenstate
\begin{equation}
q(E)=\frac{4}{N}\sum_{\alpha=1}^N\bra{E}s_\alpha^z\ket{E}^2,
\end{equation}
with this normalization $q\in[0,1]$. The average over eigenstates is
\begin{equation}
q=\frac{1}{2^N}\sum_E q(E).
\label{eq:avgqEA}
\end{equation}

To get the physical significance of this quantity, following \cite{pal2010many-body} we start with a slightly magnetized spin $\alpha$ in an infinite temperature state:
\begin{equation}
\rho_0=(\id+\epsilon s^z_\alpha)/2^N
\end{equation}
with magnetization $\langle s^z_\alpha\rangle_0=\Tr{\rho_0 s^z_\alpha}=\epsilon/4$ (as $s_z^2=1/4$). The same magnetization at large time $t$ in the diagonal approximation reads
\begin{equation}
\langle s^z_\alpha\rangle_\infty=\lim_{t\to\infty}\Tr{e^{-iHt}\rho_0e^{iHt}s^z_\alpha}= \frac{\epsilon}{2^N}\sum_E\bra{E}s^z_\alpha\ket{E}^2.
\end{equation}
Therefore, averaging over $\alpha$ we obtain the equality with eq.\ (\ref{eq:avgqEA}):
\begin{equation}
q=\frac{1}{N}\sum_{\alpha}\frac{\langle s^z_\alpha\rangle_\infty}{\langle s^z_\alpha\rangle_0},
\end{equation}
namely the previously defined EA order parameter is the average survival fraction of the initial magnetization after very long times.

We notice two more things:\cite{viola2007generalized} one, that $q$ is related to the average purity of the state (here we use the total $S^z=0$):
\begin{equation}
q=\frac{2}{N}\sum_{\alpha}\Tr{\rho_\alpha^2}-1
\end{equation}
and two, that $q$ is related to the average Hamming distance of the points in configuration space when sampled with the probability distribution $p_a$:
\begin{equation}
d_{a,b}=\sum_{\alpha=1}^N\frac{1-4\bra{a}s_\alpha^z\ket{a}\bra{b}s_\alpha^z\ket{b}}{2},
\end{equation}
and multiplying by $p_{a},\ p_{b}$ and summing over $a,b$ we find:
\begin{equation}
\label{eq:qd}
L \equiv \avg{d}=\frac{N}{2}(1-q).
\end{equation}

So $q$ is computationally easy and it captures both some geometric properties of the covering of the configuration space by an eigenstate and the long-time correlation function for $s^z$. We averaged $q$ over the spectrum (sample over typical states) and then over realizations (the number of which depends on the size of the system but it will never be less than 100).

\begin{figure}[htbp]
\centering
 \includegraphics[width=8cm]{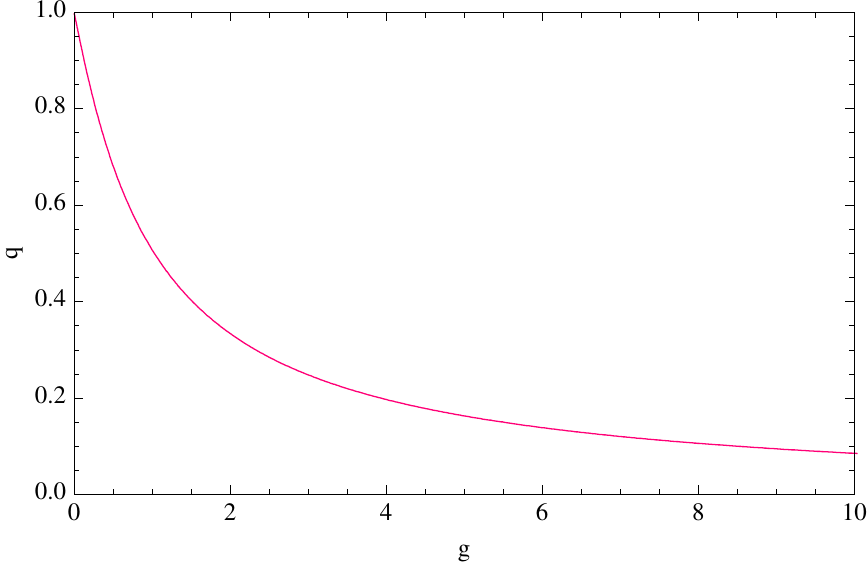}
\caption{The pointwise extrapolation of the function $q$ as a function of $g$. The fit $q=1/(1+g)$ is not distinguishable from the data.}
\label{fig:qg}
\end{figure}

We found this average $\mean{q}$ as a function of $g$ for $g\in[0,40]$ and $N=16,...,38$ and studied the pointwise finite-size scaling (in the form $q_N(g)=q(g)+c_1(g)/N+c_3(g)/N^3$) to obtain the thermodynamic limit of $q$ (see Figure \ref{fig:qg}). We fit the data using a ratio of polynomials with the condition that $q(0)=1$ and we found that averaging over the state and the realization of disorder
\begin{equation}
\label{eq:qgpade}
\mean{q}=\frac{1+3\times 10^{-8}g}{1+1.003g+0.009g^2}\simeq\frac{1}{1+g}
\end{equation}
works in the whole range of data to an error of at most $0.5\%$. We therefore conjecture this to be the correct functional form of the EA order parameter at infinite temperature.

\begin{figure}[htbp]
\centering
\includegraphics[width=8cm]{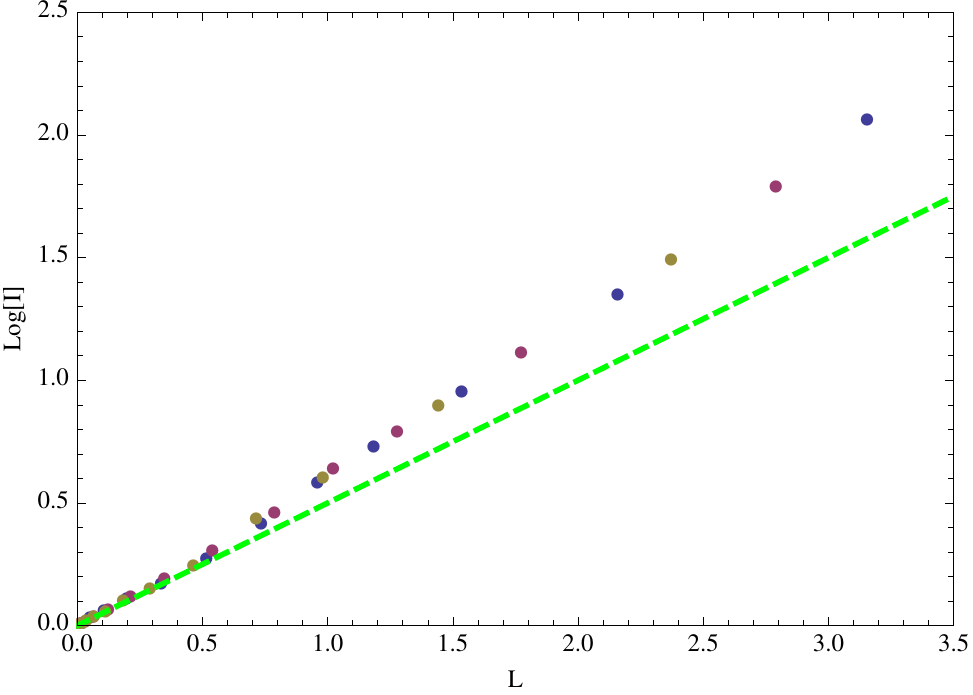}
\caption{$\log \cI$ as a function of the average distance $L$. The points are (blue, pink, yellow) $N = 28,30,32$ averaged over 100 realizations: the dashed straight line is the second order perturbation theory approximation Eq.\ (\ref{ipdrel}).}
\label{fig:IPRL}
\end{figure}

We can now go back to the relationship between the IPR and $q$, better expressed as a relation between $\log\cI$ and $L$. We notice a one-to-one correspondence between average values these two quantities already at the level of second-order perturbation theory in $g$ starting from a given state with $N/2$ spins up $\cS_\uparrow$ and $N/2$ spins down $\cS_\downarrow$:
\begin{equation}
\cI = 1 + \frac{2 g^2}{N^2} A + o(g^2)
\end{equation}
where we defined a sum over pairs of up and down spins of the given state:
\begin{equation}
A = \sum_{\alpha \in \cS_{\uparrow},\, \beta \in \cS_{\downarrow}} \frac{1}{ (h_\alpha - h_\beta)^2}
\end{equation}
Since $A$ is dominated by small denominators, it will be typically $A = O(N^4)$ and therefore from the expression for IPR we see the perturbative regime is valid for $g \ll1/N$. With an analogous computation we get:
\begin{equation}
 \label{Qperturbation}
L = \frac{4 g^2}{N^2} A + o(g^2)
\end{equation}
Eliminating $g$ between the two relations and using (\ref{eq:qd}) one obtains, independently of the state and of the quenched randomness (therefore the relation holds also on average):
\begin{equation}
\label{ipdrel}
\log \cI \simeq \frac{L}{2}.
\end{equation}
So the relation is linear for small $g$. To see how this relation is modified at higher values of $g$ we have again to resort to numerics. From the data it is clear that a strict relation exists between $\mean{\log\cI}$ and $\mean{L}$ as one can see in Figure \ref{fig:IPRL}

By using the previous Montecarlo calculation for $\cI$ we can plot $\log\cI$ vs.\ $L$, showing that the relation is almost linear. The degree of non-linearity is measured by the ratio
\begin{equation}
s=\frac{\log\cI}{2L}
\end{equation}
which can be interpreted as a local entropy\cite{campbell1987random}. In fact, $2L=N(1-q)$ can be interpreted as the number of free spins (whose value of $s^z$ is close to 0) while $\cI$ is the number of configurations. If we want $2L$ spins to be responsible to $\cI$ states then each of these spins should account for a degeneracy of $e^s$, from which the interpretation as an entropy density.

The distribution of $L$ over states and realizations becomes more and more peaked as $N$ grows, since we observe the variance $\delta L^2\propto N$. The same occurrs to $\log\cI$, whose variance goes $\propto N$ in the region of $g$ considered.  Therefore the average value of $s$ becomes typical in the large-$N$ limit.

\begin{figure}[htbp]
\centering
\includegraphics[width=8cm]{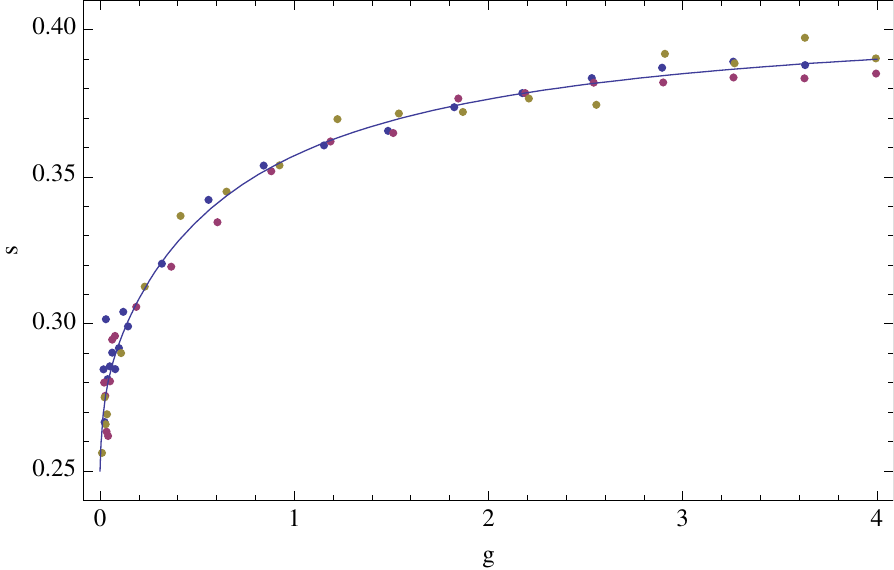}
\caption{Local entropy as a function of $g$. The points are $N=28,32,36$ (blue, pink, yellow) all averaged over 100 realizations. The fit is a (1,1)-Pade' approximation conditioned to $s(0)=1/4$.}
\label{fig:sg}
\end{figure}

\begin{figure}[htbp]
\begin{center}
\includegraphics[width=8cm]{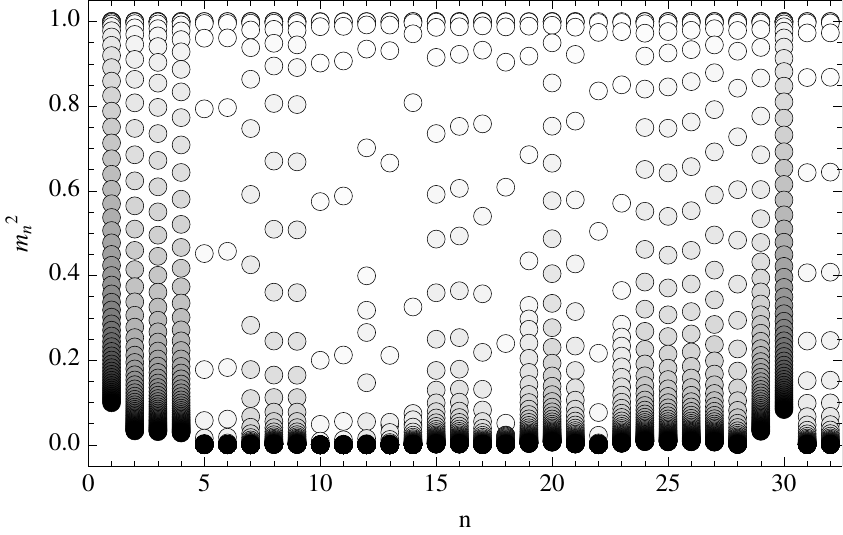}
\caption{Squared magnetizations for increasing $g$ (from white to black), where the spins are ordered by increasing magnetic field $h_n$. A consistent number of valence bond pairing is observed as a good fraction of neighboring spins (e.g. $n=8,9$, $n=15,16$ and $n=31,32$) have the same speed.}
\label{fig:magspeed}
\end{center}
\end{figure}

In the curves of Figure \ref{fig:sg} the entropy $s$ grows from the value of $1/4=0.250$ predicted by perturbation theory to an asymptotic value of $s=0.383\pm0.003$.\footnote{In the figure we show also a rational function best fit $s(g)=0.403\ \frac{0.452+g}{0.656+g},
$ which has however an error of $5\%$ in the asymptotic value $s(\infty)=0.403$ instead of $0.383$, the value obtained by averaging on many more realizations and including smaller $N$ in the fit.} This value is not what one would expect from a uniform superposition over $\binom{N}{N/2}$ states, since in that case $L=N/2$, $\log\cI\simeq N\log 2$ and the familiar value $s=\log 2=0.693$ is roughly \emph{twice} as much as we expect. This leads us to think that the most probable structure of the delocalized state at increasing $g$ still retains a pair structure. We can build a toy model of delocalization in the typical eigenstates, by assuming that $Nq$ spins are localized on their $g=0$ values and that the remaining $N(1-q)$ spins are instead divided into couples, where couples are formed between almost resonating spins of opposite orientation. The couples are in one of the random valence bond states $\ket{\uparrow\downarrow}\pm\ket{\downarrow\uparrow}$ which are indeed the two $S_z=0$ eigenstates of the two-body Hamiltonian
\begin{equation}
H_2=-\frac{g}{N}(s^+_1s^-_2+s^+_2s^-_1)-h (s^z_1+s^z_2)
\end{equation}
where we have assumed that $h_1\simeq h_2=h$.
This predicts that
\begin{equation}
\cI\sim2^{N(1-q)/2}=2^L
\end{equation}
and so that we should have a constant entropy $s=\log(2)/2=0.347$, slightly smaller than the observed value at large $g$ and off by 40\%  at small $g$. The pair structure of a given eigenstate can be observed in Figure \ref{fig:magspeed} where we plot the values of $m_n^2\equiv\bra{E}s_n^z\ket{E}^2$ for a given eigenstate $\ket{E}$, ordering the spins by increasing $h_n$ (so that almost-resonant spins are nearest neighbors). We see a clear valence bond-like pair correlation in the values of the squared magnetization.

With the available data, we can discuss issues like the presence of multiple clusters in the same energy level $\ket{E}$. In fact, by randomly restarting the Montecarlo routine with the same $p_a$'s if multiple clusters exist, we would expect to sample them according to their basin of attraction. Moreover we can rely on analytic results (such as those for $q$) to compare the Montecarlo averages with: clusterization and ergodicity-breaking would translate in a difference between these two results (as the random-walk would get stuck in a cluster and would not explore the whole configuration space). In the region where we can trust our numerics ($ g \gtrsim 0.02 $) Montecarlo averages converge to the analytic results, though a slowdown of the dynamics is observed (see below).

\subsection{Dynamics of Montecarlo and other quantities} 
The Montecarlo routine which allows for importance sampling of the distribution $p_a$ allows other measures of the geometry of the state. We can now study the similarities between the dynamics of importance sampling on $p_a$ and that of  random percolation on the hypercube, which has been proposed as a model of relaxation in a glassy system \cite{campbell1987random}. We will find that in both cases, a stretched exponential is the best fit and that the exponent depends on the coupling constant $g$. This, we believe, is a remarkable similarity.

\begin{figure}[htbp]
\centering
\includegraphics[width=8cm]{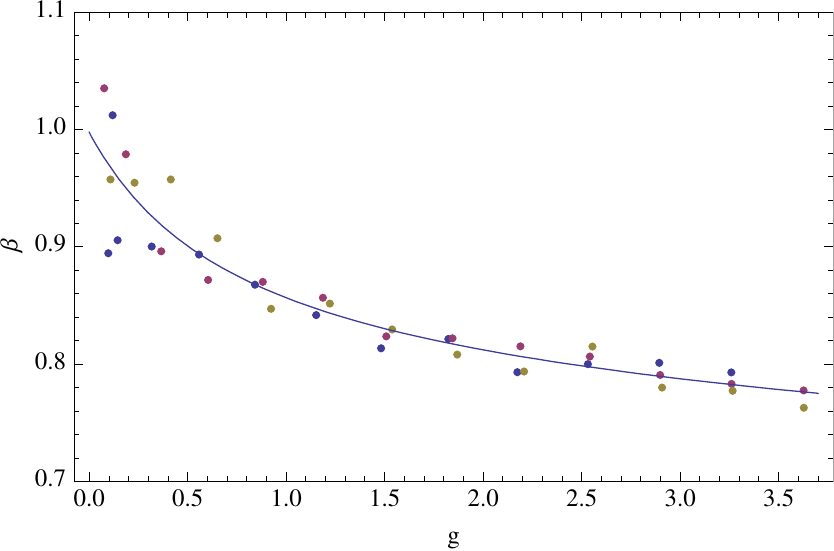}\\
\includegraphics[width=8cm]{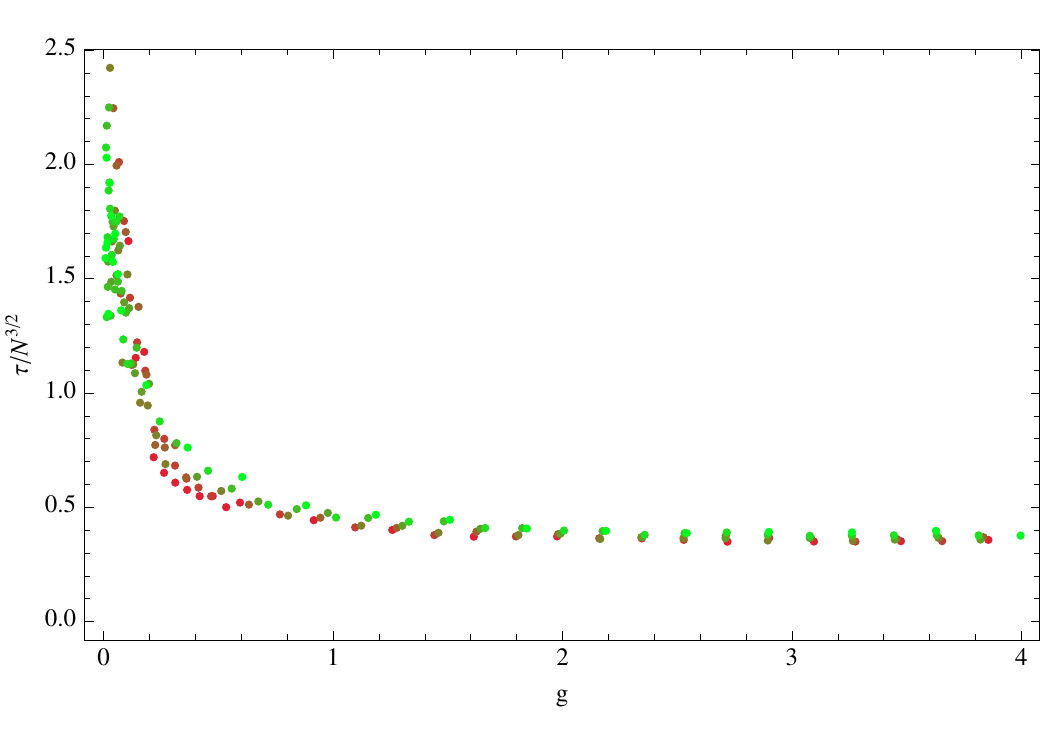}
\caption{\emph{Up.} The stretched exponential exponent $\beta$ data as a function of $g$ for $N=28,32,36$ (blue, pink, yellow) together with a fit of the form $(1+a_1 g)/(b_0+b_1 g+b_2 g^2)$. \emph{Down.} The timescale $\tau$ as a function of $g$ for $N=18,20,...,34$ (red to green). The scaling $\tau\propto N^{3/2}$ is evidently good, in particular in the region $g>1$.}
\label{fig:betag}
\end{figure}

An important quantity in this sense is the time dependence of the average distance from the starting point. Consider the Hamming distance $H(t)$ ($t$ is Montecarlo time) from the starting point $H(t)=|a(t)-a(0)|$. For $t \gg 1$, $H(t)$ is fit quite accurately by a stretched exponential ansatz of the form:
\begin{equation}
H(t) = L \left(1 - e^{-\left(\frac{t}{\tau}\right)^\beta} \right),
\end{equation}
where $L$ is the average distance introduced before and $\beta$ is a new characteristic exponent.
Let us consider the behavior of the exponent $\beta$ with respect to $g$, as plotted in Fig.\ \ref{fig:betag}. Even if the results become quite noisy for small $g$, we can still see that for small values of $g$, $\beta$ stays close to $1$, while as $g$ increases, $\beta$ decreases, although quite slowly. Instead for the time-scale $\tau$ we find, apart from the monotonic decrease with $g$, which is to be expected on general grounds that, for $g\gtrsim 1$, $\tau\propto N^{3/2}$ which we propose without explanation.

The small time behavior of $H(t)$ can be used to obtain some information about the local structure of the state. In particular we can set
\begin{equation} 
k \equiv \frac{H(1)}{2} = \frac{4}{N^2} \sum_{\langle a, b \rangle} \min (p_a, p_b) 
\end{equation}
where the last equality follows from the Montecarlo rate and the sum is over nearest-neighbour states. This quantity can be considered as a measure of the local connectivity, that is, the average fraction of active links. 

\begin{figure}[htbp]
\begin{center}
\includegraphics[width=8cm]{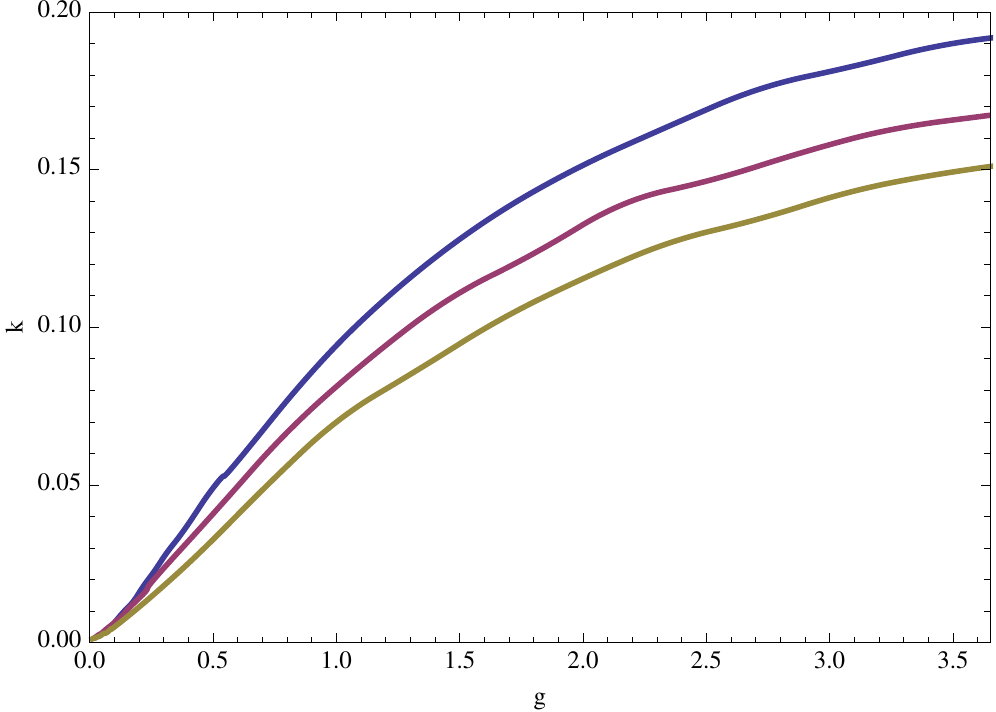}
\caption{Local connectivity as a function of $g$. Different lines corresponds to $N = 18,24,30$ (blue, pink, yellow)}
\label{fig:conng}
\end{center}
\end{figure}

From Fig.\ \ref{fig:conng}, we may deduce two things: one is that the connectivity stays well below $1$ even for large $g$, confirming, as we claimed before, that the typical state is never uniformly spread over the hypercube; the second is that the connectivity scales with $N$ as $N^{-1}$ for small $g$ and with $N^{-1/2}$ for large $g$ (a fit $k=A/N^\alpha$ shows a continuously decreasing $\alpha$ from 1 to $1/2$). Since the number of diverging roots is proportional to the total spin of the eigenstate at infinite $g$, and since the more roots diverge the more equally distributed the terms of each creation operator $B$ are, we can argue the state will be better spread for larger total spin $S$. As for large $N$, the total spin of a typical state will increase only as $\sqrt{N}$ this explains the depletion of the local connectivity. 

\subsection{Some perturbation theory}
Let us see what the predictions of perturbation theory are for our system. Let us start at $g=0$ from state $a$, with energy $E_{a}$. The states at distance 2 from $a$ have energies
\begin{equation}
\Delta^{(2)}=E_{b}-E_{a}=h_\alpha-h_\beta
\end{equation}
where the couple $(\alpha,\beta)\in \cS_\uparrow\times\cS_\downarrow$ defines the spins which have been flipped up and down in going from $a$ to $b$. The typical value of $\Delta^{(2)}$ is $\sqrt{\avg{(\Delta^{(2)})^2}}=\sqrt{2}=\Ord{1}$ however the minimum value is $\Ord{N^{-2}}$ which we write $x^{(2)}/N^2$ where $x^{(2)}=\Ord{1}$. So the corresponding term in perturbation theory for the wave function is
\begin{equation}
A_{b}=\frac{(g/N)}{x^{(2)}/N^2}=\frac{gN}{x^{(2)}}.
\end{equation}
In this way we can go on at arbitrary distance $2k$, to the state $b^k$, the amplitude thus having $k$ denominators of $\Ord{1/N^2}$
\begin{equation}
A_{b^{k}}=\frac{(gN)^k}{x^{(2)}x^{(4)}...\ x^{(2k)}},
\end{equation}
where $x$ are random variables of $\Ord{1}$. For any given $a$ there are only $\Ord{1}$ neighboring states with $\Delta\sim 1/N^2$, so the number of such $b^k$ states at distance $2k$ from $a$ is $\Ord{1}$ out of $N^{2k}$ (also the number of relevant paths does not grow as $k!$). These can be called a \emph{direct} or \emph{percolating} contribution. However already at distance 4 we observe another type of contribution, which one is tempted to dub a \emph{tunnelling} contribution, in which although the final denominator $\Delta^{(4)}=h_\alpha-h_\beta+h_\gamma-h_\delta=z^{(4)}/N^4$ each of the two paths leading to the minimum $h_\alpha-h_\beta\simeq -(h_\gamma-h_\delta)=y^{(2)}=\Ord{1}$, where $\alpha,\gamma\in\cS_\uparrow$ and $\beta,\delta\in\cS_\downarrow$. Again this contribution is of order: 
\begin{equation}
A_{b}=\frac{(g/N)}{y^{(2)}}\frac{(g/N)}{z^{(4)}/N^4}=\frac{(gN)^2}{y^{(2)}z^{(4)}},
\end{equation}
while the amplitudes corresponding to the distance-2 intermediate steps are $\Ord{1/N}$.
The distribution of $x,z$ can be found by using the theory of extreme value statistics \cite{kotz2000extreme}, while $y$'s are typical values of field differences and none of these distribution depends on $N$. We will stop here our analysis of perturbation theory as this would require a separate work by itself. It is sufficient for us to notice that only the combination $gN$ appears in all terms of the series so scaling $g$ to zero like $1/N^{1+\epsilon}$ for every $\epsilon>0$ each term would go to 0 and the series would trivially converge unless the series is asymptotic in $gN$. Notice that an argument based on a Bethe-lattice approximation for the configuration space $\cC$ (see \cite{abou1973selfconsistent}) would give $g_c\sim 1/N \log N$ for the localization transition.

\subsection{Setup of an exponential IPR} 

From perturbation theory (and from the Bethe-lattice approximation result \cite{abou1973selfconsistent}) one could argue that, if a phase transition occurs, it is at $g\sim 1/N$ (an extra factor $1/\log N$ would not be noticed for our moderately large $N$). But does a phase transition in the geometric properties of the eigenstate occur?

First we analyze the quantity for which we have more extensive statistics (because of his polynomial complexity), $L$. A phase transition in $L$ would mean that, set $\gamma=g N$, there exists a $\gamma_c>0$ such that for $\gamma<\gamma_c$, $L/N\to 0$  and for $\gamma>\gamma_c$, $L/N\sim(\gamma-\gamma_c)^\delta$ where $\delta$ is a critical exponent. We have analysed our data for $g>0.01$ and $g<0.2$ and we can conclude that this is not the behavior observed. The behavior is more consistent with $\gamma_c=0,\ \delta=1$ or with a crossover, in which the limit $L/N$ when $N\to\infty$ is a smooth function of $g$ which vanishes at $g=0$. The matching with the part of the curve at finite $g$ is smooth and the limiting behavior is as described before.

\begin{figure}[htbp]
\begin{center}
\includegraphics[width=8cm]{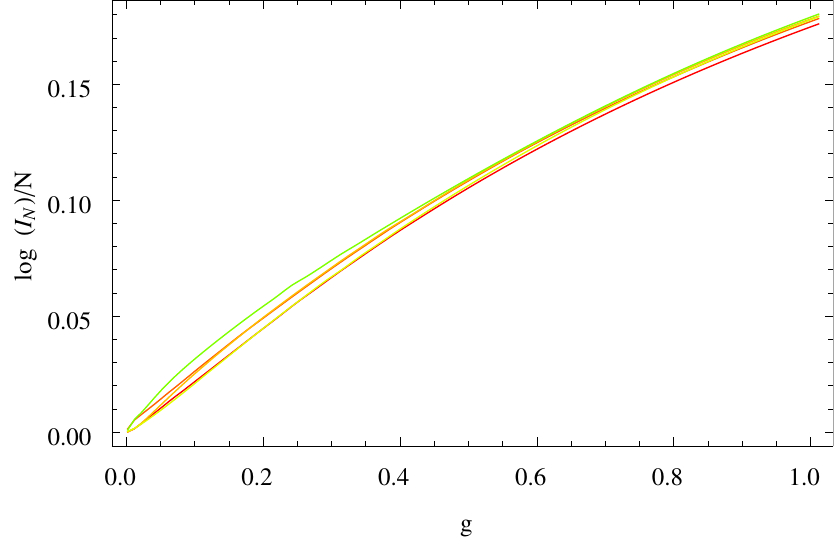}
\caption{The logarithm of the inverse participation ratio divided by the size for $N$ between $8$ and $18$, computed exactly and averaged over states and realizations. This graph shows no hint of a phase transition at $g\lesssim 1/N$.}
\label{fig:xactlogipr}
\end{center}
\end{figure}

There is the possibility however that although $L\sim N$ always, we have two phases: $\log\cI\sim 1$ and $\log\cI\sim N$ between which a transition occurs. This could happen if an eigenstate spread along one (or a few) directions without covering an exponential number of spin configurations. We have excluded this by both direct analysis of $\log\cI/N$ data and by the observation that the relation between $\log\cI$ and $L$ remains valid all the way to small $g$ (small here means $g\lesssim 1/N$). As $\log \cI/N$ becomes soon independent of $N$ without any scaling of $g$ needed (see Fig.\ \ref{fig:xactlogipr}) we are led to conclude that no phase transition occurs as the system occupies an exponential number of sites of the computational basis for any $g>0$. 

\section{Breaking of integrability}

By considering the Richardson model essentially as a hopping process on the hypercube with random site energies given by the unperturbed energies we have found that the eigenstates are always covering an exponential number of spin configurations but nonetheless $q>0$ meaning thermalization is not achieved. 
From this point of view it is not obvious which role, if any, integrability plays.

But, on the other hand, one would infer that the integrability of the model must play an essential role beyond providing the methods used for its solution. In particular, if the integrals of motion are too much local, integrability can have the effect of freezing the expectation values of local quantities. To support such a claim we investigated a very similar non-integrable model, in which the hopping coefficients are not uniformly equal to $g$ as in (\ref{hamiltonian}) but instead $N(N-1)/2$ random variables $g_{\alpha,\beta}=g(1+\epsilon\eta_{\alpha,\beta})$ where $\eta_{\alpha,\beta}=\eta_{\beta,\alpha}=\pm 1$ with probability $1/2$. Randomness in the fields is retained. The hamiltonian is
\begin{equation}
H=-\frac{g}{N}\sum_{\alpha,\beta}(1+\epsilon\eta_{\alpha,\beta})s^+_\alpha s^-_\beta-\sum_{\alpha}h_\alpha s^z_\alpha.
\label{eq:Hnoni}
\end{equation}

We observe a decrease of the value of $q$ (averaged both over $E,\eta$ and $h$) as expected, in particular for sufficiently large $g$ we are confident to say that $q\to 0$ for $N\to\infty$ and the system becomes ergodic.

\begin{figure}[htbp]
\begin{center}
\includegraphics[width=7.5cm]{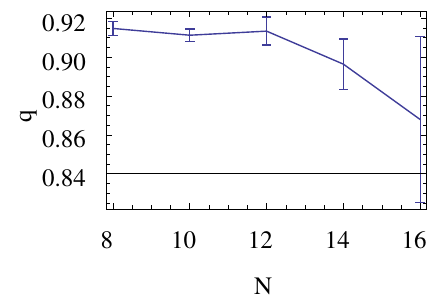}
\includegraphics[width=7.5cm]{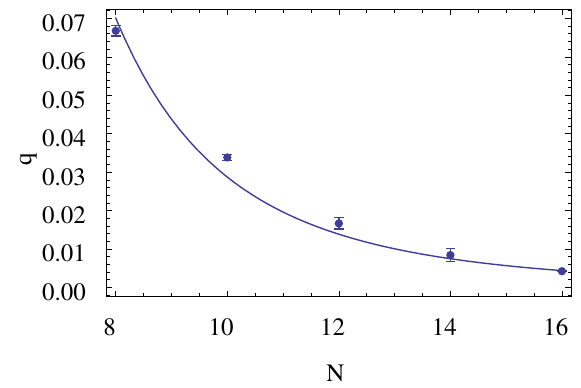}
\caption{Averaged microcanonical $q$ for the non-integrable Hamiltonian eq.\ (\ref{eq:Hnoni}) for \emph{up:} $g=0.1,\ \epsilon=0.4$ and \emph{down}: $g=4.1,\epsilon=0.4$. The data in the lower panel are fit by a power law $aN^{-\gamma}$ with $a=290$ and $\gamma\simeq 4$. The exponent $\gamma$ seems to be $g$-dependent.}
\label{fig:qnonin}
\end{center}
\end{figure}

For small $g$ however the situation is not so clear. The limit $N\to\infty$ could actually be zero or not, what is clear is that the $N$-dependence is not settled (compare the upper and lower panel of Figure \ref{fig:qnonin}) for $N=16$, the largest system size that we can attain. This leads to two competing scenarios: in the first we have ergodicity as soon as $\epsilon>0$; in the second, one could identify a finite $g_c(\epsilon)$ such that for $g<g_c$, $q>0$ and for $g>g_c$ we have $q=0$ in the thermodynamic limit. The latter would have a MBL transition at the said $g_c$. Much more extensive numerical work is needed to decide between these two scenarios. We leave the resolution of this issue for the future.

\section{Conclusions and some directions for further work} 
We have performed a numerical study of typical states of the Richardson model with quenched disorder (an example of Gaudin magnet and an integrable system). We have found no evidence of a delocalization phase transition although typical eigenstates occupy an exponential number of states in the basis of $s^z_i$'s for any $g>0$. 

We have devised a method to calculate the IPR without summing over exponentially many states and studying its connections with a microcanonical version of the Edwards-Anderson order parameter, which measures the fraction of surviving magnetization at infinite temperature and for long times. Of this order parameter, we have conjectured the thermodynamic limit at infinite temperature as $q=1/(1+g)$. We were unable to obtain the temperature dependence of this quantity, as sampling from the Boltzmann distribution is not straightforward within our framework.

For what concerns the absence of a MBL \emph{phase transition} we can point out two peculiarities of our system as responsible for its absence. One is integrability and the other is the infinite range of the Hamiltonian. We have therefore studied small-size systems (up to $N=16$ spins) with an extra integrability breaking term of size $\epsilon$. We observe a sharp reduction of $q$, which in some range of parameters could lead to think to a phase transition where $q=0$ for $g>g_c(\epsilon)$. However it is possible that in the complementary region $(g<g_c(\epsilon))$ the decrease with $N$ starts from a value of $N$ impossible to reach with our limited numerics so we are unable to see that $q=0$ for all $g$ as soon as $\epsilon>0$. 
Unfortunately this dichotomy is unlikely to be settled with the currently accessible values of $N$ and in absence of an established scaling theory of MBL.

We point out that the Richardson model is one of a family of integrable spin systems (generalized Gaudin's magnets, see \cite{ortiz2005exactly}) which can be studied with minor modifications of the methods introduced in this paper. We leave this too for further work. 

\section{Acknowledgements}
It is a pleasure for us to thank M.Mueller, G.Santoro, S.Sondhi, A.Silva, X.-Q. Yu and in particular R.Fazio and D.Huse for their suggestions and comments.

\bibliography{Richardsontypicalv2}

\end{document}